\documentstyle[12pt]{article}
\normalbaselines

\begin{document}
\bibliographystyle{unsrt}
CRM-1852
\def\ra{\rightarrow}
\def\al{\alpha}
\def\2pi{1\over 2\pi i}
\def\q{q-q^{-1}}
\def\.{\mathaccent"005F}
\def\={\mathaccent"0016}
\def\^{\mathaccent"007E}
\def\~{\tilde}
\def\newline{\hfil\break}
\def\gam{\gamma}
\def\la{\lambda}
\def\ra{\rightarrow}
\def\va{\varphi}
\def\pa{\partial}
\def\sq2{\sqrt{2}}
\def\sqk2{\sqrt{2(k+2}}
\def\sqk{\sqrt{k}}
\def\sqs{\sqrt{2\over k}}
\def\ps{\psi}
\def\psd{\psi^{\dagger}_1(z)}
\def\psdl{\psi^{\dagger}_\ell(z)}
\def\be{\begin{equation}}
\def\ee{\end{equation}}
\def\bar{\begin{array}}
\def\ear{\end{array}}
\def\bea{\begin{eqnarray}}
\def\eea{\end{eqnarray}}
\vbox{\vspace{10mm}}
\vskip 1.5truecm
\begin{center}
{\LARGE \bf Uniqueness of the bosonization of the
$U_q(su(2)_k)$ quantum current algebra}\\[10mm]
A.H. Bougourzi\\
[2mm]{\it CRM,
Universit\'e de Montr\'eal\\
C.P. 6128-A\\
Montr\'eal, P.Q., Canada, H3C 3J7}\\[5mm]
\end{center}
\\[1in]
\begin{abstract}

Four apparently different bosonizations of the $U_q(su(2)_k)$  quantum
current algebra for arbitrary level $k$ have recently been   proposed in
the literature. However, the relations among them have so far remained
unclear except in one case.  Assuming a special standard form for the
$U_q(su(2)_k)$ quantum currents, we derive a set of general consistency
equations that  must be satisfied. As particular solutions of this set of
equations, we recover two of the four bosonizations and we derive a new
and simpler one. Moreover, we show that the latter three, and the remaining
two bosonizations which cannot be derived directly  from this set of
equations since by  construction they do not  have the standard form, are all
related to each other through some redefinitions of their Heisenberg  boson
oscillators.
\end{abstract}
\newpage
\section{ Introduction}

There is presently much interest in the field of quantum groups  and  algebras
because of their rich applications in integrable  models of  quantum field
theory   and statistical systems, and
conformal field theory
\cite{Dri86,Jim86,Wor87,Alval89,Dev89,FrRe92,BeLe92,Daval92}.    Quantum
current (affine)  algebras  (QCA's) are of
special interest since they arise as both
dynamical and infinite dimensional symmetries of some physical  models, which
are therefore highly constrained. The most
standard example of these models is certainly the one-dimensional
XXZ quantum  spin chain in the thermodynamic limit  \cite{Daval92}.
However,    other examples of two-dimensional massive integrable  models of
quantum field theory like the sine-Gordon model are  also known to
possess quantum current symmetries \cite{BeLe92}.
Another important motivation for studying the QCA's
 is that one hopes that the great success of the usual
current algebras in the context of conformal field theory
\cite{KnZa84} will  carry over
also  to the quantum case without many  difficulties.

In particular, one hopes to extend the bosonization procedure
(which is also  known as the Feigin-Fuchs construction
\cite{FeFu82},  Dotsenko-Fateev construction
\cite{DoFa84}, and free field realization)
of conformal field theory to the case of QCA's.
The reason is that this bosonization procedure makes
the technical study and calculation of some relevant quantities
like the correlation functions, the irreducible characters, the   partition
functions, the BRST-like cohomology structure and the  fusion rules more
transparent and accessible. This extension, if  possible,  provides us then
with a promising alternative to the  quantum inverse scattering method
\cite{BaBe92} in  computing the  exact correlation   functions of  the
massive integrable models.

One important ingredient of the bosonization recipe is the
realization of the current algebra symmetries (or any other
infinite-dimensional  symmetries)
of the integrable models in terms of free boson fields.
This is because the free boson fields are generating functions  of boson
oscillators satisfying Heisenberg algebras, which are of  course  much
simpler to handle. For example, the correlation  functions  of the  vertex
fields (which are the  dynamical fields of an integrable model) are readily
computed    in terms of the much simpler
two-point correlation functions of the free boson fields.
This  obviously assumes that these vertex fields can  be
realized in terms of the free boson fields. This is in fact a  second
important ingredient of the bosonization recipe that we  will not consider
here.

The bosonization  of the usual ``classical"
$su(2)_k$ current algebra with its rich consequences
\cite{ItKa90,Geral90,Bou92,Neme89,Jayal90} is  well
known and extensively studied in the literature.
It depends on whether the level $k$ is equal to 1 or a generic  positive
integer number (only the unitary case is being considered). In the  former
case,
it is  called the Frenkel-Kac bosonization and  requires one free boson field
 \cite{FrKa80}. In the latter case,  after the bosonization of the pair
of ghost fields first  introduced in the Wakimoto construction, it is
referred to as the Wakimoto bosonization and requires three
boson fields   \cite{Wak86}. Then a natural question
arises   as to whether it is possible to extend this bosonization to
the $U_q(su(2)_k)$ QCA case. The answer to
this question turns out  to be positive. As expected, it
depends again on whether the level $k$ is equal to 1 or a
generic positive integer number. In the former case, it is known  as  the
Frenkel-Jing construction and requires one deformed
Heisenberg algebra \cite{FrJi88}.
In order to keep the analogy between the classical and quantum  cases
apparent, we will  occasionally use the language of the
Heisenberg algebras instead of that of the free boson fields.     The reason is
that in the quantum case several deformed  boson   fields,
which are different generating functions built from the  same Heisenberg
algebra, might be required. More details  clarifying this point   will be given
later. In the latter case,
four apparently different bosonizations have recently been
constructed independently by four different groups, namely,
Abada-Bougourzi-El Gradechi (ABE) \cite{BoGr92,Abaal92}, Matsuo
\cite{Mat92}, Shiraishi \cite{Shi92,Katal92}, and Kimura \cite{Kim92}.
All these  bosonizations require  three independent
(intercommuting) deformed Heisenberg algebras.
By apparently different, we mean that deformations of the
Heisenberg algebras, and therefore the quantum currents
 involved in one bosonization are different from those of the  other three.
However,  except for the connection between the Matsuo
and Kimura constructions \cite{Kim92}, the relations among these  bosonizations
 have  so far only been raised but not resolved in    the literature.

In this paper, we  elucidate these relations.  Strictly
speaking, we derive a
set of  general equations by assuming a  standard form for the  quantum
currents, i.e., they are expressed in terms of a  ``bosonized" parafermion
plus a free boson. Each
solution  of this set of equations leads to a
particular bosonization of the $U_q(su(2)_k)$ QCA.
In particular, we  recover the ABE and Matsuo  bosonizations,
which are by construction already  in the  standard form, and we  derive a new
much simpler  one as three different solutions of this same set of
equations. Moreover, we  show that these three bosonizations  together with
the  Shiraishi  and Kimura bosonizations, which cannot be recovered from
this set of equations since  they are not originally constructed  in the
standard form, are all
related to each other through some redefinitions of the three  sets  of
deformed Heisenberg boson oscillators.

This paper is organized as follows. In section 2, we briefly  review the
usual    $su(2)_k$  current algebra, and its
corresponding Frenkel-Kac  and Wakimoto bosonizations. In   section 3,  we
introduce the $U_q(su(2)_k)$ QCA. After
reviewing  the Frenkel-Jing bosonization in a slightly
different language for the one used by Frenkel and Jing, we examine in
detail the   bosonization of the
$su(2)_k$ QCA for a generic positive level
$k$, which we refer to as ``the deformation of the Wakimoto  bosonization."
The    analogy with the classical case is
stressed all along. The aformentioned consistency equations are  written
down.    Then, we show how the ABE and Matsuo bosonizations are    recovered
as special solutions of these equations, which in addition
lead to a new  compact bosonization that we refer to as  ``the
 fifth  bosonization." In section 4,
we elucidate the relations among all these five bosonizations.  More precisely,
we show that each of the known four bosonizations  can be obtained from the
fifth one through some linear  transformations relating their three sets of
deformed  Heisenberg boson oscillators.
This is  obviously sufficient to show that all of the five  bosonizations
can be    obtained from one another in this manner.
 Up to these transformations,
this result suggests then that the deformation of the Wakimoto
bosonization is also unique as in the classical case. We  devote
section 5 to the conclusions.

\section{ The $su(2)_k$ current algebra}

The $su(2)_k$ current algebra at level $k$  is generated by the   currents
$H(z)$ and $E^\pm(z)$ ($z$ being a complex variable),  which in the
Cartan-Weyl basis have the operator product
expansions (OPE's)
\bea
H(z).H(w)&\sim &{k\over (z-w)^2},
\label{ope1}\\
H(z).E^{\pm}(w)&\sim& \pm{\sqrt{2}E^\pm(w)\over z-w},
\label{ope2}\\
E^+(z).E^-(w)&\sim &{\sqrt{2}H(w)\over z-w}+{k\over (z-w)^2}.
\label{ope3}\eea
where the symbol $\sim$ as usual means  an equality up to
regular terms as $z$ approaches $w$.  Using standard  techniques  in
conformal field theory to re-express the OPE's of
currents (generating functions) as  commutators of their  modes  \cite{GoOl86},
 we get the more familiar $su(2)_k$ affine algebra \cite{Kac85}
\be
\bar{rcl}
{[H_n,H_m]}& = & nk\delta_{n+m,0},\\
{[H_n,E^{\pm}_m]}&=&
\pm\sqrt{2}E^\pm_{n+m},\\
{[E^+_n,E^-_m]} &=& \sqrt{2}H_{n+m}+nk\delta_{n+m,0},
\label{calg}
\ear
\ee
where
\be\bar{rcl}
H(z)&=&\sum_{n=-\infty}^{+\infty}H_nz^{-n-1},\\
E^\pm(z)&=&\sum_{n=-\infty}^{+\infty}E^\pm_nz^{-n-1}.
\label{gen}
\ear\ee

Let us now briefly review the bosonization of the
$su(2)_k$ current algebra as given by (\ref{ope1})-(\ref{ope3}). As mentioned
in   the introduction, this bosonization depends on whether the level
$k$ is  equal to 1 or a generic positive integer number, and  is  called
the Frenkel-Kac or  Wakimoto bosonization respectively.

\subsection{  The Frenkel-Kac bosonization}

In this case ($k=1$), the currents $H(z)$ and $E^\pm(z)$ are
realized  in terms of one free boson field $\phi^1(z)$  as:
\bea
H(z)&=&i\pa\phi^1(z),\\
E^\pm(z)&=&\exp\{\pm i\sq2\phi^1(z)\}.
 \eea
By free field, we mean that $\phi^1(z)$ has the OPE
$\phi^1(z).\phi^1(w)\sim -\ln(z-w)$, which is the free Green  function of the
two-dimensional Laplace equation. This OPE can be  translated  into
commutation relations if we write $\phi^1(z)$ as a generating  function,
i.e.,
\be
\phi^1(z)=\phi^1-i\phi^1_0\ln{z}+i\sum_{n\neq 0}{\phi^1_n
\over n}z^{-n},
\ee
in which case the set of boson oscillators
$\{\phi^1,\phi^1_n\}$ satisfies  the Heisenberg algebra
\be\bar{rcl}
{[\phi^1_n,\phi^1_m]}&=&n\delta_{n+m,0},\\
  {[\phi^1,\phi^1_0]}&=&i.
\label{FKH}
 \ear\ee
In fact, this is valid only if we define the normal ordering of  the boson
oscillators,  which is  denoted by the symbol ::,    such that
\be\bar{rcl}
: \phi^1_n\phi^1_m:&=& \phi^1_m\phi^1_n, \qquad n>0,\\
: \phi^1_0\phi^1:&=& \phi^1\phi^1_0.
\label{nor}
\ear\ee
The symbol :: is understood and therefore omitted from any field  or product
of fields defined at the same point $z$. These  definitions and  conventions
will
be valid throughout the rest of  this paper.

\subsection{  The Wakimoto bosonization}

This is again valid for a generic positive level $k$.
Though the Wakimoto construction is based on one free boson field  and a
pair of ghost fields \cite{Wak86}, it is well known that it is  equivalent to
the
following construction which involves only
the free boson fields $\phi^1(z)$, $\phi^2(z)$ and $\phi^3(z)$:
\bea
H(z)&=&i\sqk\pa\phi^1(z),
\label{H}\\
E^\pm(z)&=&\left(\pm i\sqrt{{k\over 2}}\pa\phi^2(z)+
i\sqrt{k+2\over 2}
\pa\phi^3(z)\right)\exp\{\pm  i\sqs(\phi^2(z)+\phi^1(z))\},
\label{E}
\eea
where $\phi^1(z)$ is the same as the one used in
the Frenkel-Kac bosonization. These three free boson fields are   orthogonal to
each other in the sense that if we write them as  generating functions
\be
\phi^j(z)=\phi^j-i\phi^j_0\ln{z}+i\sum_{n\neq 0}
{\phi^j_n\over  n}
z^{-n},\qquad j=1,2,3,
\label{cphi}
\ee
the corresponding boson oscillators satisfy the three
intercommuting Heisenberg algebras
\be\bar{rcl}
{[\phi^j_n,\phi^\ell_m]}&=&(-1)^{j-1}n\delta^{j\ell}
\delta_{n+m,0},\\
{[\phi^j,\phi^\ell_0]}&=&(-1)^{j-1}i\delta^{j\ell},\qquad\qquad  j=1,2,3,
\label{qkWH}
\ear\ee
with all the other commutators being trivial. Note the presence  of an extra
minus sign for the set of oscillators $\{\phi^2,\phi^2_n\}$.
It is  primarily the above form of the currents $E^\pm(z)$ that we refer
to as the standard form,  i.e., a ``bosonized" parafermion plus a free boson.
 This is because $E^\pm(z)$ can be rewritten  as \cite{ZaFa85}:   \be
E^\pm(z)=\sqk\psi^\pm(z)\exp\{\pm i\sqs\phi^1(z)\},
\ee
where  $\phi^1(z)$ is a free boson field and
\be
\psi^\pm(z)=\left(\pm i\sqrt{{1\over 2}}\pa\phi^2(z)+
i\sqrt{k+2\over 2k}
\pa\phi^3(z)\right)\exp\{\pm i\sqs\phi^2(z)\}
\ee
are parafermions since they  satisfy the OPE
\be
\psi^+(z).\psi^-(w)\sim (z-w)^{2/k}.
\label{pf}
\ee

This standard form amounts essentially to  realizing the current  $H(z)$ as a
derivative of a  single free boson field
$\phi^1(z)$ as shown in (\ref{H}). Note that if  we  do not include  the
parafermion in $E^\pm(z)$, we will obtain  an almost trivial  generalization
(just a   normalization factor $\sqk$ will arise) of the Frenkel-Kac
bosonization.
This  is because the OPE's (\ref{ope1}) and (\ref{ope2}) will be already
satisfied.
However, the remaining OPE (\ref{ope3})  is not satisfied and  no freedom is
left over with only one free boson field $\phi^1(z)$ to  satisfy  it.
Consequently, we need
to add a  parafermion, which is in turn ``bosonized"  in terms of  two new
independent free boson fields $\phi^2(z)$ and  $\phi^3(z)$ as in (\ref{E}).
Note that the inclusion of this parafermion does not spoil the fact that the
currents  $E^\pm(z)$ still  satisfy the  OPE's (\ref{ope1}) and  (\ref{ope2}).
The reason is that the current $H(z)$ depends only  on  $\phi^1(z)$ that is
orthogonal to  $\phi^2(z)$ and $\phi^3(z)$,  which give rise to
 this  parafermion. These remarks will be important when extending the
Frenkel-Jing bosonization ($k=1$) to the case of a generic  positive
level $k$. In  fact, both the Frenkel-Jing and Wakimoto bosonizations will
serve as a guiding  tool to derive  their respective quantum analogues.

\section{The $U_q(su(2)_k)$ quantum current algebra}

In order to underline the analogy with the classical case, let us  first
introduce this algebra in the language of commutation relations. It is
generated by the operators
$\{H_n, E^\pm_n,\quad n\in {\bf Z}\}$ and
reads in the Cartan-Weyl basis as \cite{Abaal92,Dri85,Jim85}:
\be\bar{rcl}
&&{[H_n,H_m]} = {[2n][nk]\over 2n}\delta_{n+m,0},\qquad
n\neq 0,\\
&&{[H_0, H_m]}=  0,\\ && {[H_n,E^{\pm}_m]}=
\pm\sqrt{2}{q^{\mp |n|k/2}[2n]\over 2n}
E^\pm_{n+m}, \q
quad n\neq 0,\\
&&{[H_0, E^\pm_m]} =  \pm\sq2 E^\pm_m,\\
&&{[E^+_n,E^-_m]} = {q^{k(n-m)/2}\Psi_{n+m}-q^{k(m-n)/2}
\Phi_{n+m}\over q-q^{-1}},\\
&&E^\pm_{n+1}E^\pm_m-q^{\pm 2}E^\pm_mE^\pm_{n+1}=
q^{\pm 2}E^\pm
_nE^\pm_{m+1}-E^\pm_{m+1}E^\pm_n,
\label{qalg}
\ear\ee
where the familiar notation $[x]=(q^x-q^{-x})/(q-q^{-1})$
is used, $q$ is the deformation parameter which is not a
root of  unity here, and $\Psi_n$ and $\Phi_n$ are the modes of the fields
$\Psi(z)$ and $\Phi(z)$ that are defined by
\be\bar{rcl}
\Psi(z)&=&\sum_{n\geq 0}\Psi_nz^{-n}=q^{\sqrt{2}H_0}
 \exp\{\sqrt{2}(\q)\sum_{n>0}H_nz^{-n}\},\\
\Phi(z)&=&\sum
_{n\leq 0}\Phi_nz^{-n}=q^{-\sqrt{2}H_0}
\exp\{-\sqrt{2}(\q)\sum_{n<0}H_nz^{-n}\}.
\ear\ee
One can easily verify that as $q$ approaches 1
this quantum algebra (\ref{qalg}) reduces to the classical one   (\ref{calg}).
For the sake of bosonization, it is convenient to rewrite this
quantum algebra as OPE's. Now the role of the current     $H(z)$ in the
classical case (\ref{H}) will be played instead by
the currents $\Psi(z)$ and $\Phi(z)$. The $U_q(su(2)_k)$ QCA
then reads  \cite{Abaal92,Mat92,Ber89}
\bea
\Psi(z).\Phi(w)&=&
{(z-wq^{2+k})(z-w^{-2-k})\over (z-wq^{2-k})(z-w^{-2+k})}
\Phi(w).\Psi(z),
\label{ope4}\\
\Psi(z).E^{\pm}(w)&=&
q^{\pm 2}{(z-wq^{\mp(2+k/2)})\over z-wq^{\pm (2-k/2)}}
E^\pm(w).\Psi(z),
\label{ope5}\\
\Phi(z).E^{\pm}(w)&=&
q^{\pm 2}{(z-wq^{\mp(2-k/2)})\over z-wq^{\pm (2+k/2)}}
E^\pm(w).\Phi(z),
\label{ope6}\\
E^+(z).E^-(w)&\sim& {1
\over w(\q)}\left\{{\Psi(wq^{k/2})\over z-wq^k}-
{\Phi(wq^{-k/2})\over z-wq^{-k}}\right\},
\label{ope7}\\
E^{\pm}(z).E^{\pm}(w)&=&{(z q^{\pm 2}-w)\over z-w q^{\pm 2}}
E^{\pm}(w). E^{\pm}(z),
\label{ope8}
\eea
where $E^\pm(z)$ are generating functions of the modes $E^\pm_n$  as in
(\ref{gen}) and the symbol $\sim$  means that the regular terms  as $z$
approaches    $wq^{\pm k}$ are
being omitted. To go from (\ref{qalg}) to  (\ref{ope4})-(\ref{ope8})
and vice versa the identification $\sum_{n\geq 0}z^n=
(1-z)^{-1}$ with $|z|<1$ is used. This will also be understood in  the
subsequent  development. Let us now first review in this notation the
quantum analogue
of the Frenkel-Kac bosonization, namely, the Frenkel-Jing  bosonization.

\subsection{The Frenkel-Jing bosonization}

This bosonization is again valid only for $k=1$. The currents
$\Psi(z)$, $\Phi(z)$
and $E^\pm(z)$, which satisfy the OPE's (\ref{ope4})-(\ref{ope8})
are now realized as follows \cite{FrJi88,Abaal92}:
\bea
\Psi(z)&= &\exp\left\{ i\sq2\left(\va^{1,+}
(zq^{1/2})
-\va^{1,-}(zq^{-1/2})
\right)\right\}\label{psi}\\
&=& q^{\sqrt{2}\va^1_0}
\exp\{\sqrt{2}(\q)\sum_{n>0}\va^1_nz^{-n}\},
\nonumber  \\
\Phi(z)&= &\exp\left\{  i\sq2\left(\va^{1,+}
(zq^{-1/2})-
\va^{1,-}(zq^{1/2})\right)
\right\}\label{phi} \\
&=&
q^{-\sqrt{2}\va^1_0}
 \exp\{-\sqrt{2}(\q)\sum_{n<0}\va^1_nz^{-n}\},
 \nonumber\\
E^\pm(z)&=&\exp\left\{\pm i\sq2\va^{1,\pm}
(z)\right\}.
\label{qE}
\eea
Here $\va^{1,\pm}(z)$ are two different deformations of the same  free boson
field $\phi^1(z)$ introduced in the Frenkel-Kac  bosonization,   that is,
they are two different generating functions of the boson  oscillators
$\va^1_n$ and $\va^1$ satisfying the same deformation of the Heisenberg
algebra used in the Frenkel-Kac bosonization (\ref{FKH}). This means they
both  reduce to $\phi^1(z)$ as $q$ tends to 1. In this regard, $\va^1_n$ and
$\va^1$ are deformations (as suggested by the symbols) of
  $\phi^1_n$ and $\phi^1$ respectively.  More specifically,  the
deformed free fields $\va^{1,\pm}(z)$ are given by
\be
\va^{1,\pm}(z)=\va^1-i\va^1_0\ln{z}
+i\sum_{n\neq 0}{q^{\mp |n|/2}\over [n]}\va^1_nz^{-n},
 \ee
with the deformed Heisenberg algebra being
\be\bar{rcl}
{[\va^1_n,\va^1_m]}&=&{[2n][n]\over 2n}\delta_{n+m,0},\\
{[\va^1,\va^1_0]}&=&i.
\ear\ee
It can easily be checked with the normal ordering defined in
(\ref{nor}) that the quantum currents $\Psi(z)$, $\Phi(z)$ and
$E^\pm  (z)$ as realized through (\ref{psi})-(\ref{qE}) do indeed satisfy the
$U_q(su(2)_1)$ QCA (\ref{ope4})-(\ref{ope8}).

\subsection{ Deformation of the Wakimoto bosonization}

Let us now consider the extension of the previous bosonization to  the
$U_q(su(2)_k)$ QCA case in a way which is parallel as much as possible to the
manner  the Wakimoto bosonization was extended from the Frenkel-Kac
one in the  classical case. In particular, we will  maintain the standard form
for
the  quantum currents.    Consequently, the generalization of the currents
$\Psi(z)$, $\Phi(z)$  and  $E^\pm(z)$ (\ref{psi})-(\ref{qE})
so that the OPE's (\ref{ope4})-(\ref{ope6}) are satisfied is given by
\bea
\Psi(z)&=&\exp\{
i\sqs\left(\va^{1,+}(zq^{k/2})-\va^{1,-}(zq^{-k/2})
\right)\}\label{kpsi}\\
&=&
q^{\sqrt{2k}\va^1_0}
\exp\{\sqrt{2k}(\q)\sum_{n>0}\va^1_nz^{-n}\},
\nonumber\\
\Phi(z)&= &\exp\{
i\sqs\left(\va^{1,+}(zq^{-k/2})-\va^{1,-}(zq^{k/2})\right)
\}\label{kphi}\\
&=&
q^{-\sqrt{2k}\va^1_0}
\exp\{-\sqrt{2k}(\q)\sum_{n<0}\va^1_nz^{-n}\},
\nonumber\\
E^\pm(z)&=&\exp\{\pm i\sqs\va^{1,\pm}(z)\},
\label{qkE}
\eea
where now we have
\be
\va^{1,\pm}(z)=\va^1-i\va^1_0\ln{z}
+ik\sum_{n\neq 0}{q^{\mp |n|k/2}\over [nk]}\va^1_nz^{-n},
\ee
and
\be\bar{rcl}
{[\va^1_n,\va^1_m]}&=&nI_1(n)\delta_{n+m,0},\\
{[\va^1,\va^1_0]}&=&i.
\label{qWH1}
\ear\ee
Here $I_1(n)={[2n][nk]\over 2kn^2}$. The usefulness of the
notation $I_1(n)$
will be clear shortly. By analogy to the  classical case as
explained below
 equation (\ref{pf}), this achieves  fully the bosonization  of the currents
$\Psi(z)$ and
$\Phi(z)$,  but only partially that of the currents $E^\pm(z)$,   that is,
their
$\va^{1,\pm}(z)$ exponential parts. To see this, $E^\pm(z)$ as   given  by
(\ref{qkE}) satisfy
the following OPE's instead of the one given in (\ref{ope7}):
\be\bar{rcl}
E^\pm (z).E^\mp (w)&=& \exp \left\{{2\over k}\langle
\va^{1,\pm}(z)\va^{1,\mp}(w)\rangle\right\}:
E^\pm(z).E^\mp (w):,\\
E^\pm (z).E^\pm (w)&=& \exp \left\{-{2\over k}\langle
\va^{1,\pm}(z)\va^{1,\pm}(w)\rangle\right\}:E^\pm(z).E^\pm (w):,
\ear\ee

where the various two-point correlation functions that result
from the normal ordering of the boson oscillators are
given by
\be\bar{rcl}
\langle \va^{1,\pm}(z)\va^{1,\mp}(w)\rangle
&=&-\ln z+{k\over 2}\sum_{n>0}
{[2n]\over n[nk]}z^{-n}w^n,\\
\langle \va^{1,\pm}(z)\va^{1,\pm}(w)\rangle&=&
-\ln z+{k\over 2}\sum_{n>0}{q^{\mp nk}[2n]\over
n[nk]}z^{-n}w^n.
\ear\ee

To  derive a correct and complete  bosonization of $E^\pm(z)$ that is
consistent  with (\ref{ope7}) we need to introduce two more deformed free
boson fields $\va^2(z)$ and $\va^3(z)$ or rather two more sets of
deformed boson oscillators $\{\va^2, \va^2_n\}$ and
$\{\va^3, \va^3_n\}$ since we do not yet know   how the former  are
generating functions of the latter.  The sets $\{\va^2, \va^2_n\}$ and
$\{\va^3, \va^3_n\}$ are respectively the deformations (quantum   analogues)
of  the sets  $\{\phi^2, \phi^2_n\}$ and $\{\phi^3, \phi^3_n\}$
introduced in the Wakimoto  bosonization. They make up a
deformed parafermion. The main  purpose of this paper is to  determine the
deformation of the two Heisenberg algebras generated by $\{\va^2, \va^2_n\}$
and  $\{\va^3,\va^3_n\}$ and use them to complete the  bosonization of
$E^\pm(z)$. It is mainly this part that distinguishes from one  another
the several  bosonizations that have been recently
proposed in the literature.  The Wakimoto bosonization of  $E^\pm(z)$
involves classical derivatives as shown in (\ref{E}) and therefore
its natural deformation  must also involve some   quantum   derivatives.
Furthermore, it must also coincide with the Wakimoto  bosonization
(\ref{H})-(\ref{E}) or any of its equivalent forms through some  field
redefinitions as $q$ tends to 1. Finally, it must produce the two different
simple poles $z=wq^{\pm k}$ in the OPE  $E^+(z).E^-(w)$. Guided
with these constraints, the general form of the bosonization     of the quantum
currents $E^\pm(z)$ reduces to
\be
E^{\pm}(z)={\exp\{\pm i\sqs\va^{1,\pm}(z)\}\over z(\q)}
\left(\exp\{\pm i\sqs X_A^\pm(z)\}-\exp\{\pm i\sqs X_B^\pm(z)\}   \right),
\ee
where the $\va^{1,\pm}(z)$ exponential part is already fixed by  (\ref{qkE})
and
\be
X_A^\pm(z)=
\va^2-i\va^2_0\ln{zq^{A^\pm_2}}-i\va^3_0\ln{q^{A^\pm_3}}+
i\sum_{n\neq 0}\{A^\pm_2(n)
\va^2_n+A^\pm_3(n)
\va^3_n)\} {z^{-n}\over n}.
\label{XA}
\ee
$X_B^\pm(z)$ is given by a similar expression to (\ref{XA}) with  $A$ being
replaced by $B$. The boson oscillators $\{\va^2,  \va^2_n\}$
and $\{\va^3,  \va^3_n\}$ satisfy the
following deformed Heisenberg algebras:
\be\bar{rcl}
{[\va^j_n,\va^\ell_m]}&=&(-1)^{j-1}nI_j(n)\delta^{j,\ell}
\delta_{n+m,0},\\
{[\va^j,\va^\ell_0]}&=&(-1)^{j-1}i\delta^{j,\ell},
\qquad\qquad\qquad\qquad  j,\ell=2,3.
\ear\ee
Here no sum with respect to $j$ is meant. The question of the  deformation
of the Wakimoto bosonization translates now into fixing  the
unknown parameters $A^\pm_2(n)$, $A^\pm_2$, $A^\pm_3(n)$,  $A^\pm_3$,
$B^\pm_2(n)$, $B^\pm_2$,
$B^\pm_3(n)$,  $B^\pm_3$, $I_2(n)$ and $I_3(n)$
consistently with (\ref{ope7}) and the above constraints.

First, the  consistency with (\ref{ope7}) requires the following  relations
to  be satisfied:
\be\bar{rcl}
\exp\{i\sqs X^+_B(z)\}.\exp\{-i\sqs X^-_A(w)\}&=&
{z-wq^{k+2}\over q(z-wq^k)}\exp \left\{-{2\over k}\langle
\va^{1,+}(z)\va^{1,-}(w)\rangle\right\}\\
&&\times :\exp\{i\sqs \left(X^+_B(z)-X^-_A(w)\right)\}:,\nonumber\\
X^+_B(wq^k)&=&X^-_A(w),\\
\exp\{i\sqs X^+_A(z)\}.\exp\{-i\sqs X_B^-(w)\}&=&
{q(z-wq^{-k-2})\over z-wq^{-k}}\exp \left\{-{2\over k}\langle
\va^{1,+}(z)
\va^{1,-}(w)\rangle\right\}\\
&&\times:\exp\{i\sqs  \left(X_A^+(z)-X_B^-(w)\right)\}:,\nonumber\\
X_A^+(wq^{-k})&=&X_B^-(w),\\
\exp\{i\sqs X_A^+(z)\}.\exp\{-i\sqs X_A^-(w)\}&=&
q\exp \left\{-{2\over k}\langle
\va^{1,+}(z)\va^{1,-}(w)\rangle\right\}\\
&&\times:
\exp\{i\sqs \left(X_A^+(z)- X_A^-(w)\right)\}:,\nonumber\\
\exp\{i\sqs X_B^+(z)\}.\exp\{-i\sqs X_B^-(w)\}&=&
q^{-1}\exp \left\{-{2\over k}\langle
\va^{1,+}(z)\va^{1,-}(w)\rangle\right\}\\
&&\times
:\exp\{i\sqs \left(X_B^+(z)- X_B^-(w)\right)\}:.\nonumber
\ear\ee
These relations translate to these  conditions  on the unknown   parameters:
\bea
A^\pm_2(n)&=&q^{-nk}B^\mp_2(n),
\label{re1}\\
A^\pm_3(n)&=&q^{-nk}B^\mp_3(n),
\label{re2}\\
A^\pm_2&=&-B^\pm_2=k/2,
\label{re3}\\
A^\pm_3&=&B^\mp_3=\pm{\sqrt{k(k+2)}\over 2},
\label{re4}
\eea
and
\be\bar{rcl}
A^\pm_2(n)A^\pm_2(-n)I_2(n)-A^\pm_3(n)A^\pm_3(-n)I_3(n)
&=&{k\over 2}\left({[n(k+2)]\over [nk]}-1\right), \quad n>0,\\
A^\pm_2(n)A^\mp_2(-n)I_2(n)-A^\pm_3(n)A^\mp_3(-n)I_3(n)&=&
{k\over 2}{[2n]\over [nk]},\quad n>0.
\label{mas}
\ear\ee
The  relations (\ref{re3}) and (\ref{re4})
are in fact derived only from the consistency with the  limit
$q\rightarrow 1$. The parameters $A^\pm_2$, $A^\pm_3$,  $B^\pm_2$  and
$B^\pm_3$ are then completely fixed by the  latter relations.    Therefore only
the parameters $A^\pm_2(n)$,  $A^\pm_3(n)$,  $B^\pm_2(n)$,
$B^\pm_3(n)$, $I_2(n)$ and $I_3(n)$ are left to be  determined. Furthermore,
since $B^\pm_2(n)$ and
$B^\pm_3(n)$ are related to $A^\pm_2(n)$ and $A^\pm_3(n)$
through (\ref{re1}) and (\ref{re2}),
 we will henceforth  focus just on $A^\pm_2(n)$, $A^\pm_3(n)$,     $I_2(n)$
and $I_3(n)$. These are restricted to satisfy the set of general  ``master"
equations (\ref{mas}). Each solution of these  equations yields
a  particular bosonization. In fact, this is how  we will now  recover   both
the
bosonizations of ABE and Matsuo,  and derive a new one called ``the fifth"
which also has  the standard form defined previously.

\subsubsection{ The ABE bosonization}

This bosonization corresponds to the following  choice  of the  parameters
$A^\pm_2(n)$, $A^\pm_3(n)$, $I_2(n)$ and
$I_3(n)$,  which do indeed satisfy  the general equations  (\ref{mas}):
\be\bar{rcl}
A^\pm_2(n)&=&q^{-nk/2},\\
A^\pm_3(n)&=&\pm {1\over 2}
\sqrt{{k+2\over k}}(q^{-nk}-1),\\
I_2(n)&=&{k\over 4}{[n(k+2)]+[2n]-[nk]\over [nk]},\\
I_3(n)&=&{k^2\over k+2}{[n][n(k+2)/2]\over [nk][nk/2]}.
\ear\ee

In order to distinguish later the ABE bosonization from the other  ones
let us introduce the following notations,
which are specific just for  this bosonization:
\be\bar{rcl}
\va^i_n&\equiv &\xi^i_n, \\
\va^i&\equiv&\xi^i,\qquad\qquad i=1,2,3.
\ear\ee
The ABE bosonization can then be summarized in this notation as
\cite{Abaal92}:
\be\bar{rcl}
\!\!\!\!\Psi(z)&= &\exp\left\{
i\sqs\left(\xi^{1,+}(zq^{k/2})-\xi^{1,-}(zq^{-k/2})
\right)\right\}\\
&=&
q^{\sqrt{2k}\xi^1_0}\exp\left\{
\sqrt{2k}(\q)\sum_{n>0}\xi^1_nz^{-n}\right\},\\
\!\!\!\!\Phi(z)&= &\exp\left\{ i\sqs\left(\xi^{1,+}(zq^{-k/2})-
\xi^{1,-}(zq^{k/2})\right)\right\}\\
&=&q^{-\sqrt{2k}\xi^1_0}
\exp\left\{-\sqrt{2k}(\q)\sum_{n<0}\xi^1_nz^{-n}\right\},\\
\!\!\!\!E^{\pm}(z)&=&{\exp\{\pm i\sqs\xi^{1,\pm}(z)\}
\over z(\q)}  \left(\exp\{\pm i\sqs\xi^2(zq^{k/2})+
i\sqrt{k+2\over 2k^2}(\xi^3(zq^k)-\xi^3(z))\}\right.\\
&&\left.-\exp\{\pm i\sqs\xi^2(zq^{-k/2})+i\sqrt{k+2\over 2k^2}
(\xi^3(zq^{-k})-\xi^3(z))\}\right),
\label{ABE}
\ear\ee
where the deformed free bosons $\xi^{1,\pm}(z)$, $\xi^2(z)$ and
$\xi^3(z)$ are given by
\be\bar{rcl}
\xi^{1,\pm}(z)&=&\xi^1-i\xi^1_0\ln{z}
+ik\sum_{n\neq 0}{q^{\mp |n|k/2}\over [nk]}\xi^1_nz^{-n},\\
\xi^j(z)&=&\xi^j-i\xi^j_0\ln z+i\sum_{n\neq 0}
{z^{-n}\over n}\xi^j_n,\qquad j=2,3.
\ear\ee
They are given as various deformed generating functions of the  boson
oscillators  $\{\xi^j, \xi^j_n,\quad j=1,2,3\}$
which satisfy the three deformed Heisenberg  algebras
\be\bar{rcl}
{[\xi^j_n,\xi^\ell_m]}&=&(-1)^{j-1}nI_j(n)\delta^{j,\ell}
\delta_{n+m,0},\\
{[\xi^j,\xi^\ell_0]}&=&(-1)^{j-1}i\delta^{j,\ell}
\qquad  j,\ell=1,2,3,
\ear\ee
with no sum with respect to $j$ being meant, and
\be\bar{rcl}
I_1(n)&=&{[2n][nk]\over 2kn^2},\\
I_2(n)&=&{k\over 4}{[n(k+2)]+[2n]-[nk]\over [nk]},\\
I_3(n)&=&{k^2\over k+2}{[n][n(k+2)/2]\over [nk][nk/2]}.
\ear\ee
As a check, note that as $q$ approaches 1 all three $I_j(n)$,  $j=1,2,3$
tend to 1, which is consistent with (\ref{qkWH}) in the  classical case, and
the
deformed free bosons $\xi^{1,\pm}(z)$,  $\xi^2(z)$ and $\xi^3(z)$ reduce to
$\phi^1(z)$, $\phi^2(z)$ and  $\phi^3(z)$ (\ref{cphi}) respectively.
Furthermore, in this same limit the quantum currents $E^\pm(z)$ and
$\{\Psi(z)-\Phi(z)\}/\sq2(q-q^{-1})$ tend to their classical  analogues
(\ref{E}) and (\ref{H}) respectively. Finally, note that the remaining OPE
(\ref{ope8}) of the $U_q(su(2)_k)$ QCA is automatically satisfied.

\subsubsection{The Matsuo bosonization}

This bosonization is characterized by the following choice of
the parameters $A^\pm_2(n)$, $A^\pm_3(n)$, $I_2(n)$ and $I_3(n)$,
which also satisfy the general equations (\ref{mas}):
\be\bar{rcl}
A^\eta_2(n)&=&{nk\over [nk]}q^{-nk/2},\\
A^{-\eta}_2(n)&=&{nkq^{-nk/2}\over [2n]}\left(
{[n(k+2)]\over [nk]}-1\right),\\
A^\eta_3(n)&=&0,\\
A^{-\eta}_3(n)&=&\eta\sqrt{k(k+2)}nq^{-nk/2}(q-q^{-1}){
[n]\over [2n]}, \\
I_2(n)&=&{[2n][nk]\over 2kn^2},\\
I_3(n)&=&{[2n][n(k+2)]\over 2n^2(k+2)},
\label{Mat1}
\ear\ee
where $\eta$ is equal to $+$ or $-$ depending on whether
$n>0$ or $n<0$ respectively.
Again as a check, notice that $I_2(n)$ and $I_3(n)$ approach 1 as  $q$ tends
to 1. In fact,
the original bosonization of Matsuo uses  slightly  different  normalization
factors from those in the above solution. To   recover the   Matsuo
bosonization in its initial  notations, which will then  distinguish it  from
 the
others in what follows, let us make the identifications\newpage
\be\bar{rcl}
\va^1_n&\equiv&\alpha_n/\sqrt{2k},\\
\va^2_n&\equiv&\=\alpha_n/\sqrt{2k},\\
\va^3_n&\equiv&\beta_n/\sqrt{2(k+2)},\\
\va^1&\equiv &-i\sqrt{2k}\alpha,\\
\va^2&\equiv &-i\sqrt{2k}\=\alpha,\\
\va^3&\equiv &-i\sqrt{2(k+2)}\beta,
\ear\ee
which lead to the commutation relations
\be\bar{rcl}
{[\al_n,\al_m]}&=&{[2n][nk]\over n}\delta_{n+m,0},\\
{[\=\al_n,\=\al_m]}&=&-{[2n][nk]\over n}\delta_{n+m,0},\\
{[\beta_n, \beta_m]}&=&{[2n][n(k+2)]\over n}\delta_{n+m,0},\\
{[\al_0,\al]}&=&1,\\{[\=\al_0,\=\al]}&=&-1,\\
{[\beta_0,\beta]}&=&1.
\ear\ee
These are the precisely the three Heisenberg algebras used
in the Matsuo bosonization, which is summarized as follows
\cite{Mat92}\footnote{ The currents $E^\pm(z)$ are in fact equal to those of
Matsuo up  to an overall minus sign, which is irrelevant because it just
reflects a $Z_2$   automorphism of the $U_q(su(2)_k)$ QCA.}:
\be\bar{rcl}
\Psi(z)&=&q^{\al_0}
\exp\{(\q)\sum_{n>0}\al_nz^{-n}\},\\
\Phi(z)&=&q^{-\al_0}
\exp\{-(\q)\sum_{n<0}\al_nz^{-n}\},\\
E^\pm(z)&=&{Y^\pm(z)\over z(\q)}
\left\{Z_\mp(zq^{k+2\over 2})W_\mp(zq^{k\over 2})^{\pm 1}-
Z_\pm(zq^{-{k+2\over 2}})W_\pm(zq^{-{k\over 2}})^{\pm 1}\right\},  \label{Mat2}
\ear\ee
where
\be\bar{rcl}
Y^\pm(z)&=&\exp\{\pm 2(\al+\=\al)\pm{(\al_0+\=\al_0)\over k}
\ln{z}
\mp\sum_{n\neq 0}{q^{\mp |n|k/2}\over
[nk]}(\al_n+\=\al_n)z^{-n}\},\\
Z_\pm(z)&=&q^{\mp\=\al_0/2}\exp\{\mp(\q)\sum_{n>0}{[n]
\over [2n]}
\=\al_{\pm n}z^{\mp n}\},\\
W_\pm(z)&=&q^{\mp\beta_0/2}\exp\{\mp(\q)\sum_{n>0}{[n]
\over [2n]}
\beta_{\pm n}z^{\mp n}\}.
\ear\ee
It can also be verified that  as $q$ approaches 1 we recover the classical
bosonization (\ref{H}) and (\ref{E})   with or
without the same normalization factors if we use  (\ref{Mat1}) or  (\ref{Mat2})
 respectively.

\subsubsection{ The fifth bosonization}

This new fifth bosonization is specified by the following   parameters,
which also satisfy the general equations (\ref{mas}):
\be\bar{rcl}
A^\pm_2(n)&=&\sqrt{{k+2\over 2}}{nk\over [nk]}q^{n(\pm   1-2)k/2},
\qquad\qquad n>0,\\
A^\pm_2(n)&=&\sqrt{{k+2\over 2}}{nk\over [nk]}
q^{\pm nk/2},\qquad\qquad n<0,\\
A^+_3(n)&=&-\sqrt{2k}{n\over [2n]}q^{n(f-k/2)},\\
A^-_3(n)&=&-\sqrt{2k}{n\over [2n]}q^{n(f-2-3k/2)},\\
I_2(n)&=&{[nk][n(k+2)]\over n^2k(k+2)}q^{nk},\\
I_3(n)&=&{[2n]^2\over 4n^2}.
\ear\ee
Here $f$ is a free parameter whose usefulness will become clear  shortly.
Again, in order to distinguish the fifth bosonization  from the  other
ones and to write it in a more compact form, let us  specify it by
the following  identifications:
\be\bar{rcl}
\va^i_n&\equiv& \chi^i_n, \qquad n\neq 0,\qquad i=1,2,3,\\
\va^1_0&\equiv& \chi^1_0,\\
\va^2_0&\equiv& \sqrt{{2+k\over 2}}\chi^2_0-
\sqrt{{k\over 2}}\chi^3_0,\\
\va^3_0&\equiv& -\sqrt{{k\over 2}}\chi^2_0+
\sqrt{{2+k\over 2}}\chi^3_0,\\
\va^1&\equiv& \chi^1,\\
\va^2&\equiv& \sqrt{{2+k\over 2}}\chi^2-
\sqrt{{k\over 2}}\chi^3,\\
\va^3&\equiv& -\sqrt{{k\over 2}}\chi^2+
\sqrt{{2+k\over 2}}\chi^3.
\label{chi}
\ear\ee
In this notation, this fifth bosonization reads simply as:
\be\bar{rcl}
\!\!\!\!\Psi(z)&= &\exp\left\{
i\sqs\left(\chi^{1,+}(zq^{k/2})-\chi^{1,-}(zq^{-k/2})\right)
\right\}\\
&=&q^{\sqrt{2k}\chi^1_0}
\exp\left\{\sqrt{2k}(\q)\sum_{n>0}\chi^1_nz^{-n}\right\},\\
\!\!\!\!\Phi(z)&= &\exp\left\{
i\sqs\left(\chi^{1,+}(zq^{-k/2})-\chi^{1,-}(zq^{k/2})\right)
\right\}\\
&=&q^{-\sqrt{2k}\chi^1_0}
\exp\left\{-\sqrt{2k}(\q)\sum_{n<0}\chi^1_nz^{-n}\right\},\\
\!\!\!\!E^{+}(z)&=&{\exp\{i\sqs\chi^{1,+}(z)+i\sqrt{{2+k\over k}}
\chi^2(z)\}\over z(\q)}
\left(\exp\{-i\chi^3(zq^{-1})\}
-\exp\{- i\chi^3(zq)\}\right),\\
\!\!\!\!E^{-}(z)&=&{\exp\{- i\sqs\chi^{1,-}(z)
\}\over z(\q)}\left(\exp\{-i\sqrt{{2+k\over k}}
\chi^2(zq^k)+i\chi^3(zq^{1+k})\}\right.\\
&&\left.-\exp\{-i\sqrt{{2+k\over k}}
\chi^2(zq^{-k})+ i\chi^3(zq^{-1-k})\}\right),\nonumber
\label{fif}
\ear\ee
where the deformed free boson fields $\chi^{1,\pm}(z)$, $\chi^2(z)$  and
$\chi^3(z)$  are given  by
\be\bar{rcl}
\chi^{1,\pm}(z)&=&\chi^1-i\chi^1_0\ln{z}
+ik\sum_{n\neq 0}{q^{\mp |n|k/2}\over [nk]}\chi^1_nz^{-n},\\
\chi^{2}(z)&=&
\chi^2-i\chi^2_0\ln{z}+ik\sum_{n\neq 0}
{q^{- |n|k/2}\over [nk]}\chi^2_nz^{-n},\\
\chi^{3}(z)&=&\chi^3-i\chi^3_0\ln{z}+2i\sum_{n\neq 0}
{q^{n(f-1-k/2)}\over [2n]}\chi^3_nz^{-n}.
\ear\ee
These fields are given as various deformed generating functions   of the  boson
oscillators which satisfy the three deformed Heisenberg  algebras  \be\bar{rcl}
{[\chi^j_n,\chi^\ell_m]}&=&(-1)^{j-1}nI_j(n)\delta^{j,\ell}
\delta_{n+m,0},\\
 {[\chi^j,\chi^\ell_0]}&=&(-1)^{j-1}i\delta^{j,\ell},
\qquad\qquad\qquad  j,\ell=1,2,3,
\ear\ee
with
\be\bar{rcl}
I_1(n)&=&{[2n][nk]\over 2kn^2},\\
I_2(n)&=&{[nk][n(2+k)]\over n^2k(2+k)}q^{nk},\\
I_3(n)&=&{[2n]^2\over 4n^2}.
\ear\ee
Note that  this bosonization not only satisfies the QCA but is   also given in
a very simple compact form. In fact, it is the  simplest of all the
bosonizations considered in this paper.

\section{ Relations among all the bosonizations of the
$U_q(su(2)_k)$ QCA}

In this section, we unravel the relations among the previous  three
bosonizations  and those of both Shiraishi and Kimura.
As we will see shortly, the  latter two      cannot be recovered directly from
the  general equations (\ref{mas}) because by construction they are not in  the
standard form. In fact, we will only show that the fifth  bosonization is  at
the
center of all  these  bosonizations in the sense that it  can be  related
to each of the other four through some redefinitions
of the three sets of deformed boson oscillators. This means that  in this way
they are all related to each other.

\subsection{ The relation between the ABE and the fifth  bosonizations}

One can easily check that the ABE bosonization (\ref{ABE}) can be  obtained
from  the fifth one (\ref{fif}) if their boson  oscillators are  related
through the following linear transformations:
\be\bar{rcl}
\xi^1_{\pm n}&=&\chi^1_{\pm n},\qquad n>0,\\
\xi^2_{\pm n}&=&{nk\sqrt{2+k}q^{-nk/2}\over 2\sq2 [nk/2]}
\chi^2_{\pm n}-{n\sqk(1+q^{\mp(2+k)})q^{\pm nf}\over
\sq2[2n]}\chi^3_{\pm n},\qquad n>0,\\
\xi^3_{\pm n}&=&-{nk\sqk q^{-nk/2}\over \sq2 [nk]}
\chi^2_{\pm n}+
{nk\sq2[n(2+k)/2]q^{\pm n(f-1-k/2)}\over
\sqrt{2+k}[2n][nk/2]}\chi^3_{\pm n},\qquad n>0,
\ear\ee
and $\{\xi^j_0, \xi^j, \quad j=1,2,3\}$ are related to
$\{\chi^j_0, \chi^j, \quad j=1,2,3\}$ in the same way as in
(\ref{chi}) with $\{\va^j_0, \va^j, \quad j=1,2,3\}$ being  substituted
by $\{\xi^j_0, \xi^j, \quad j=1,2,3\}$.

\subsection{ The relation between the Matsuo and the fifth  bosonizations}

These two bosonizations can also be related to each other through  these
transformations:
\be\bar{rcl}
\al_{\pm n}&=&\sqrt{2k}\chi^1_{\pm n},\qquad n>0,\\
\=\al_{\pm n}&=&\sqrt{k(2+k)}\chi^2_{\pm n}-{2[nk]q^{n(\pm
f+(2+k)(1\mp  1)/2)}
\over [2n]}\chi^3_{\pm n},\qquad n>0,\\
\beta_{\pm n}&=&-\sqrt{k(2+k)}q^n\chi^2_{\pm n}\mp
{2[n(2+k)]q^{n(\pm(f-1)+
(k\mp k)/2)}\over [2n]}\chi^3_{\pm n},\qquad n>0,
\ear\ee
and
\be\bar{rcl}
\al_0&=&\sqrt{2k}\chi^1_0,\\
\=\al_0&=&\sqrt{k(2+k)}\chi^2_0-k\chi^3_0,\\
\beta_0&=&-\sqrt{2(2+k)}\chi^2_0+(2+k)\chi^3_0,\\
\al&=&{i\over \sqrt{2k}}\chi^1,\\
\=\al&=&{i\over 2}{ \sqrt{2+k\over k}}\chi^2-{i\over 2}\chi^3,\\
\beta&=&-{i\over 2}{ \sqrt{k\over 2+k}}\chi^2+{i\over 2}\chi^3.
\ear\ee

\subsection{The relation between the  Kimura-Shiraishi and the  fifth
bosonizations}

In this section, we have considered simultaneously the Shiraishi   and the
Kimura bosonizations because as we  will show later they  can be treated on the
same basis as two  special  cases of a
more general single bosonization, which we refer to as the
Kimura-Shiraishi  bosonization and  define by
\be\bar{rcl}
\Psi(z)&= &q^{L_0+M_0}
\exp\{(\q)\sum_{n>0}(L_n+M_n)z^{-n}\},\\
\Phi(z)&= &q^{-(L_0+M_0)}
\exp\{-(\q)\sum_{n<0}(q^{-ng}L_n+q^{-nh}M_n)z^{-n}\},\\
E^{\pm}(z)&=&{1\over z(\q)}
\left(\exp\{M^\pm(z)+N^\pm(z)\}-
\exp\{\~L^\pm(z)+\~M^\pm(z)+\~N^\pm(z)\}\right),
\label{KS}
\ear\ee
with
\be\bar{rcl}
M^+(z)&=&\~M^+(z)=-{M\over 2}-{M_0\over 2}\ln{z}+\sum_{n<0}
{q^{  -n(h+2+k/2)}
\over [2n]}
 M_nz^{-n}+\sum_{n>0}{q^{ n(2+k/2)}\over
[2n]}M_nz^{-n},\\
N^+(z)&=&\~N^+(zq^{-2})=-{N\over 2}-{N_0\over  2}\ln{zq^{-1}}+
\sum_{n\neq 0}{q^{ n(f-k/2)}
\over [2n]}
 N_nz^{-n},\\
L^-(z)&=&L_0\ln{q}+(\q)\sum_{n>0}q^{ -nk/2}
L_nz^{-n},\\
\~L^-(z)&=&-L_0\ln{q}-(\q)\sum_{n<0}q^{n(k/2-g)}
L_nz^{-n},\\
M^-(z)&=&\~M^-(zq^{2k+4})={M\over 2}+{M_0\over 2}
\ln{zq^{2+k}}-\sum_{n<0}{q^{ -n(h+2+3k/2)}
\over [2n]}
 M_nz^{-n}-\sum_{n>0}{q^{ -n(2+k/2)}\over
[2n]}M_nz^{-n},\\
N^-(z)&=&\~N^-(zq^{2k+2})={N\over 2}+{N_0\over  2}\ln{zq^{1+k}}+
\sum_{n\neq 0}{q^{ n(f-2-3k/2)}
\over [2n]}
 N_nz^{-n}.
\ear\ee
The  sets of boson oscillators $\{ L,L_n  \}$, $\{ M,M_n\}$ and $\{ N,N_n\}$
satisfy the  three deformed Heisenberg algebras
\be\bar{rcl}
{[L_n,L_m]}&=&{[2n][n(2+k)]\over n}q^{-n(2+g)}\delta_{n+m,0},\\
{[M_n,M_m]}&=&
-{[2n]^2\over n}q^{-n(h+2+k)}\delta_{n+m,0},\\
{[N_n, N_m]}&=&{[2n]^2\over
n}\delta_{n+m,0},\\
{[L_0,L]}&=&2(2+k),\\
{[M_0,M]}&=&-4,\\
{[N_0,N]}&=&4.
\ear\ee
Here $f$ is the same free parameter as that introduced in
(\ref{fif}), and $g$ and $h$ are two new free parameters. It can  be
checked that  despite the presence of these three free   parameters, this
Kimura-Shiraishi bosonization which is not in the standard form  does indeed
satisfy the $U_q(su(2)_k)$ QCA (\ref{ope4})-(\ref{ope8}).

The fifth bosonization can also be obtained from the
Kimura-Shiraishi  one through the following redefinitions  of their respective
boson oscillators:\newpage
\be\bar{rcl}
L_n&=&{\sqrt{2k}[n(2+k)]q^{-2n}\over [nk]}\chi^1_n+
{\sqrt{k(2+k)}[2n]q^{-n(2+k)}\over [nk]}\chi^2_n,\\
M_n&=&-{\sqrt{2k}[2n]q^{-n(2+k)}\over [nk]}\chi^1_n-
{\sqrt{k(2+k)}[2n]q^{-n(2+k)}\over [nk]}\chi^2_n,\\
C_n&=&2\chi^3_n,\\
L_{-n}&=&{\sqrt{2k}[n(2+k)]q^{-n(2+g)}\over [nk]}\chi^1_{-n}+
{\sqrt{k(2+k)}[2n]q^{-n(2+g+k)}\over [nk]}\chi^2_{-n},\\
M_{-n}&=&-{\sqrt{2k}[2n]q^{-n(2+h+k)}\over [nk]}
\chi^1_{-n}-
{\sqrt{k(2+k)}[2n]q^{-n(2+h+k)}\over [nk]}\chi^2_{-n},\\
C_{-n}&=&2\chi^3_{-n},
\ear\ee
with $n> 0$, and
\be\bar{rcl}
\sqrt{{k\over 2}}L_0&=&(2+k)\chi^1_0+\sqrt{2(2+k)}\chi^2_0,\\
\sqrt{{k\over 2}}M_0&=&-2\chi^1_0-\sqrt{2(2+k)}\chi^2_0,\\
\sqrt{{k\over 2}}N_0&=&\sqrt{2k}\chi^3_0.
\label{KSf}
\ear\ee
The set $\sqrt{{k\over 2}}\{L,M,N\}$ is related to
$i\{\chi^1,\chi^2,\chi^3\}$ in a similar manner as in  (\ref{KSf}).

Let us now show how both the Shiraishi and Kimura bosonizations  can  be
recovered as two special cases from the Kimura-Shiraishi  one.  If we fix
the three free parameters as:
\be\bar{rcl}
f&=&3+3k/2,\\
g&=&-4,\\
h&=&-2k-4,
\ear\ee
and make the following identifications:
\be\bar{rcl}
L_n&\equiv& a_n, \qquad n\neq 0,\\
M_n&\equiv& b_n, \qquad n\neq 0,\\
N_n&\equiv& c_n, \qquad n\neq 0,\\
L_0&\equiv& \~a_0,\\
M_0&\equiv& \~b_0,\\
N_0&\equiv& \~c_0,\\
L&\equiv& Q_a,\\
M&\equiv& Q_b,\\
N&\equiv& Q_c,
\ear\ee
we recover  the Shiraishi bosonization in its original  notations,
that is, \cite{Shi92}
\be\bar{rcl}
\!\!\!\!\Psi(z)&=&q^{(\~a_0+\~b_0)}
\exp\{(\q)\sum_{n>0}(a_n+b_n)z^{-n}\},\\
\!\!\!\! \Phi(z)&=&q^{-(\~a_0+\~b_0)}
\exp\{-(\q)\sum_{n<0}(q^{4n}a_n+q^{n(2k+4)}b_n)z^{-n}\},\\
\!\!\!\!  E^{+}(z)&=&{q^{-(\~b_0+\~c_0)(k+2)/2}
\exp\{-b(2|zq^{-k-2};-k/2)\}\over z(\q)}
\left(\exp\{-c(2|zq^{-k-3};0)\}-
\exp\{-c(2|zq^{-k-1};0)\}\right),\\
\!\!\!\!E^-(z)&=&q^{(\~b_0+\~c_0)(k+2)/2}
{\exp\{-a(k+2|zq^{-2};
k/2)\}\over z(\q)}\\
&&\times\left(\exp\{a(k+2|zq^{k};-2-k/2)+
b(2|z;-2-k/2)+c(2|zq^{-1};0)\}\right.\\
&&\left.-\exp\{a(k+2|zq^{-k-4};-2-k/2)+b(2|zq^{-2k-4};-2-k/4)+
c(2|zq^{-2k-3};0)\}\right).\nonumber
\ear\ee
Here the following deformed boson fields have been used:
\be
d(x|z;y)={Q_d\over x}+{\~d_0\over x}\ln{z}
-\sum_{n\neq 0}{q^{|n|y}\over
[nx]}d_nz^{-n},\qquad d=a,b,c.
\ee
Note that the extra term   $q^{\pm(\~b_0+\~c_0)(k+2)/2}$, which arises here
 but not in the Shiraishi bosonization, is irrelevant in the sense that
it can be kept or omitted without affecting the correctness of this
bosonization. This is because it commutes with the currents $E^\pm(z)$,
$\Psi(z)$ and $\Phi(z)$, and moreover it does not affect their limits as $q$
approaches 1.

Similarly, the Kimura bosonization is obtained
from the Kimura-Shiraishi one if we set the three free parameters  to
\be\bar{rcl}
f&=&k+2,\\
g&=&-2,\\
h&=&-k-2,
\ear\ee
and make the identifications
\be\bar{rcl}
L_n&\equiv& b_n, \\
M_n&\equiv& a_n, \\
N_n&\equiv &\=a_n, \\
L&\equiv& b,\\
M&\equiv& a,\\
N&\equiv &\=a.
\ear\ee
We  indeed retrieve in this way  the Kimura bosonization as first
proposed in \cite{Kim92}, i.e.,\newpage
\be\bar{rcl}
\!\!\!\!\Psi(z)&=&q^{(a_0+b_0)}
 \exp\{(\q)\sum_{n>0}(a_n+b_n)z^{-n}\},\\
\!\!\!\!\Phi(z)&=&q^{-(a_0+b_0)}
\exp\{-(\q)\sum_{n<0}(q^{n(k+2)}a_n+q^{2n}b_n)z^{-n}\},\\
\!\!\!\!E^+(z)&=&{Y^+(z)\over z(\q)}
\{Z_-(zq^{k+2\over 2})-
Z_+(zq^{-{k+2\over 2}})\},\\
\!\!\!\!E^-(z)&=&{Y^-(z)\over z(\q)}
\{Z_+(zq^{k+2\over 2})W_+(zq^{k\over 2})U_+(zq^{k\over 2})-
Z_-(zq^{-{k+2\over 2}})W_-(zq^{-{k\over 2}})
U_-(zq^{-{k\over 2}})\},
\ear\ee
where
\be\bar{rcl}
\!\!\!\!Y^\pm(z)&\!\!\!\!=\!\!\!\!&\exp\{\mp {(a+\=a)\over 2}
\mp{(a_0+\=a_0)\over  2}\ln{z}\pm
\sum_{n<0}{q^{\pm nk/2}
\over [2n]}
 (a_n+\=a_n)z^{-n}\pm\sum_{n>0}{q^{n(2+k\mp k/2)}\over
[2n]}(a_n+\=a_n)z^{-n}\},\\
\!\!\!\!Z_\pm(z)&\!\!\!\!=\!\!\!\!
&q^{\mp\=a_0/2}\exp\{\mp(\q)\sum_{n>0}{[n]\over [2n]}
q^{-n(k+2)(1\mp 1)/2}\=a_{\pm n}z^{\mp n}\},\\
\!\!\!\!W_\pm(z)&\!\!\!\!=\!\!\!\!&q^{\pm b_0}\exp\{\pm(\q)
\sum_{n>0}q^{n(-1\pm1)}
b_{\pm n}z^{\mp n}\},\\
\!\!\!\!U_\pm(z)&\!\!\!\!=\!\!\!\!&q^{\pm  (k+2)(a_0+\=a_0)/2}
\exp\{\pm(\q)\sum_{n>0}
q^{nk(1\pm 1)/2}  q^{-n(1\mp 1)}{[n(k+2)]\over [2n]}
(a_{\pm n}+\=a_{\pm n})z^{\mp n}\}.
\ear\ee

\section{ Conclusions}

In this paper, we have proven that all of the four presently
available and  apparently different bosonizations of the  $U_q(su(2)_k)$
quantum current algebra are in fact equivalent to  each other and  to a
new simpler one through the redefinitions of  the Heisenberg boson
oscillators. Clearly, this result suggests  that the deformation of the
Wakimoto
bosonization with three sets  of Heisenberg boson oscillators is  also unique
up to  the  redefinitions of these oscillators as in the  classical case.
Now that the question of the bosonization of the quantum currents  which
generate the $U_q(su(2)_k)$ quantum current algebra is      clarified, the
focus should be put on the other two important     ingredients of the
bosonization recipe,  namely, the realization   of the representations of this
algebra or in the language of     conformal field theory the quantum analogue
of the primary  fields, and the realization of the quantum analogue of the
screening currents in terms of the same Heisenberg boson
oscillators.  The latter two ingredients are necessary for the
calculation of many  relevant quantities and in particular the
correlation functions of the XXZ model. Moreover, we expect the fifth
bosonization to  simplify  particularly the bosonization  of the  quantum
analogue of the primary fields. We think  that the quantum
analogue of the parafermion algebra whose elementary generators can
easily be deduced from the currents $E^\pm(z)$ of the
$U_q(su(2)_k)$ quantum current algebra also deserves  more attention.
These and other  related questions are
presently under investigation.

\section{Acknowledgments}
 We appreciate the warm hospitality of the ``Centre de Recherches
Math\'ematiques (CRM)" where this work was achieved. We also thank Dr. R.
Weston for his interesting comments on this manuscript.

\end{document}